# Nanofriction on Sodium Polystyrene Sulfonate Brushes in Water

T. Fujima*, E. Futakuchi and F. Kino

Department of Mechanical Engineering, Tokyo City University,

1-28-1 Tamazutsumi, Setagaya, 158-8557 Tokyo, Japan.

*Corresponding author: E-mail: tfujima@tcu.ac.jp.

We investigated the frictional properties of sodium polystyrene sulfonate (NaPSS) brushes in water by frictional force microscopy (FFM). Polyelectrolyte brushes were prepared on silicon wafers by the "grafting to" method. The brushes considerably reduce the frictional force and coefficient of kinetic friction compared to hydrodynamic lubrication on a smooth Si wafer. Frictional force is independent of sliding speed, but is lower for lower degrees of NaPSS polymerization. Nanoindentation tests indicate that the polymer chains in a brush are stretched strongly away from the substrate. These results suggest that polymer chains point support the FFM probe tip in water and reduced contact area and friction.

INTRODUCTION

A polymer brush[1] is often fabricated on a solid surface by one of two chemical methods: "grafting to" or "grafting from." Brushes so fabricated have particularly superior properties of heat resistance, solvent resistance, and mechanical strength compared with brushes fabricated by physical methods such as the Langmuir–Blodgett method. Polyelectrolyte brushes (PEBs)[2,3] are brushes that have dissociable group; they form monolayer films on substrates and are promising as two-dimensional nanoactuators,[4] high-efficiency matrices for catalytic metal particles,[5] lubricants for artificial joints[4] and for nano-technologies.[6]

The frictional properties of PEBs in solvent are particularly interesting. Klein and coworkers showed that uncharged polymer brushes reduce frictional force in a good solvent by osmotic pressure, and PEBs do the same by electrostatic interactions, as determined by frictional force measurements performed using a surface force apparatus.[7-9] Zhao and coworkers showed on a micrometer scale that uncharged polystyrene brushes reduce friction in toluene.[10] Bielecki and coworkers also did for oil-compatible polymer brushes.[11] Kobayashi and coworkers showed on a millimeter scale that amphipathic polymer brushes reduce the coefficient of kinetic friction.[12]

Most studies of the frictional properties of PEBs have been performed on PEBs that contain not strong but rather weak polyelectrolytes such as poly(methyl methacrylate). A frictional study on a strong PEB, sodium polystyrene sulfonate (NaPSS) brush, was conducted by Ohsedo and co-workers in macroscopic scale by rheometer[13]. Studies about nano-friction on PEBs that contain strong polyelectrolytes should provide new insight into the mechanism of low-friction PEB behavior. In this study, we investigated the nanometer-scale frictional properties of PEBs that contain NaPSS in water, including the dependence of frictional force on load, sliding speed, and polyelectrolyte chain length.

SAMPLES

NaPSS with mono-dispersed molecular weight distribution was used as a strong polyelectrolyte to form PEBs. PEB samples were fabricated on Si wafers (thickness: 525 mm) by the method reported by Tran and Auroy.[14] $SiCl_3$-end-capped polystyrene was prepared by anion polymerization in benzene (molecular weight dispersion: 1.07). The substrates were cleaned and hydroxylated by ultrasonication and vacuum-ultraviolet–ozone exposure, immersed in a toluene solution of the $SiCl_3$-end-capped polystyrene for 1 h, and heated in $N_2$ at 160 °C for 1 day to give uncharged polystyrene (PS) brushes where the PS chains had one of two degrees of polymerization: $n$ = 55 or 110. The PS brushes were first sulfonated in a mixture of acetic anhydride and sulfuric acid diluted with 1,2-dichloroethane and then immersed in an $NaHCO_3$ aqueous solution to exchange the counterions and give the desired NaPSS brushes.

The graft density $D$ of the brushes was determined before sulfonation by ellipsometry. $D$ is defined by the relationship $D = h\,d\,N_A\,M_w^{-1}$, where $h$ is the dry thickness of the PS layer given by ellipsometry, $d$ is weight density, $N_A$ is Avogadro's number, and $M_w$ is molecular weight. For the weight density of dried PS, we used $d$ = 1.06 g/cm$^3$. The graft density of the PS brush so obtained was converted into the graft density of the sulfonated NaPSS brush by including the degrafting rate of the polymer chains, derived by infrared absorption measurements as described below.

The degree of sulfonation and degrafting of the brush chains during sulfonation were determined by Fourier-transform infrared spectroscopy (Spectrum One, Perkin Elmer) performed before and after sulfonation. The spectra show bands at 2924 and 3026 cm$^{-1}$ ascribed to asymmetric $CH_2$ stretching and monosubstituted phenyl C–H stretching, respectively. The intensity of the former is proportional to the number of alkyl segments, while that of the latter is proportional to the number of unsulfonated styrene units. Thus, comparing the spectra before and after sulfonation is useful for our purposes. The measured sulfonation rate for all samples and all degrees of polymerization was 0.8. The graft densities of the NaPSS brushes were 0.51 chains/nm$^2$ ($n$ = 55) and 0.26 chains/nm$^2$ ($n$ = 110).

EXPERIMENTS

Nanoscale frictional properties were investigated by frictional force microscopy (FFM) using a scanning probe microscope (SPA400, SII NanoTechnology) and a silicon cantilever (NSC36/AlBS, MikroMasch; bending spring constant: 1.75 N/m; probe tip radius: 10 nm). The sample and cantilever were set up under purified water. Frictional force was obtained from the torsion spring constant $k$ of the cantilever, calculated from its bending spring constant and dimensions by the equation

$$k = \frac{2wt^3G}{3l(2h+t)} \quad (1)$$

where $k$ is the coefficient of torsion spring [Nm/rad]; $w$, $t$, $l$, and $h$ are the width, thickness, length, and height of the cantilever [m], and G is the shear modulus [Pa].

Kinetic frictional force was measured as a function of both test load and sliding speed. For load-dependence measurements, sliding speed was fixed at 0.8 $\mu$m/s and the test load was varied from 1.75 to 175 nN. For sliding-speed-dependence measurements, the test load was fixed at 17.5 nN and the sliding speed was varied from 20.0 to 62.5 $\mu$m/s. All friction measurements were performed in a scanning area of 300 × 300 nm; only the central area (100 × 100 nm) was averaged to avoid the effect of static friction.

Force curve was measured in water by nanoindentation tests (TI-900 TriboIndenter, Hysitron) with a Berkovich-type diamond indenter designed for measurements in liquid (tip curvature radius: 100 nm; angle between edges: 115°). The load velocity was fixed at 0.04 $\mu$N/s.

RESULTS AND DISCUSSION

Figure 1 shows plots of kinetic nanofrictional force as a function of test load for NaPSS brushes ($n$ = 55 and 110) and a bare Si wafer for comparison. For all samples, frictional force increases linearly with increasing test load. Although the frictional force on Si wafer follows the Amontons–Coulomb law[15], being proportional to test load without offset friction, the frictional force on the NaPSS brushes does show obvious offset friction at zero load. Mean-square analysis of the frictional properties from the figure gives the coefficients of kinetic friction listed in Table 1. The value for Si wafer is 0.008, which is typical for fluid lubrication. The values for the NaPSS brushes are considerably lower.

Figure 2 shows plots of kinetic nanofrictional force as a function of sliding speed for NaPSS brushes ($n$ = 55 and 110). Frictional force is clearly independent of sliding speed, suggesting that the friction mechanism on NaPSS brushes is not fluid lubrication. In addition, the viscous resistance from ambient water and grafted polymer chains is negligible because the resistance should be proportional to sliding speed. If the probe tip had penetrated the brush layer, it would have encountered viscous resistance from the surrounding polymer chains. Therefore, this result indicates that the probe was supported on the free end of the strongly stretched NaPSS chains.

Figure 3 shows plots of load as a function of displacement depth obtained during nanoindentation tests of NaPSS brushes ($n$ = 110). As the indenter penetrates the brush, load increases and changes in gradient to a depth of 20 nm, comparable to the length of an $n$ = 110 polymer chain. Thus, the shallow region (<20 nm) corresponds to the brush layer and the deep region corresponds to the substrate.

The gradient of the load-displacement curve for the brush layer—that is, its elasticity—is about one-third of that for the substrate, indicating that the polyelectrolyte chains of the brush are stretched strongly away from the substrate. A condition of such strong stretching, called an "osmotic region," is known to form when strong osmotic pressure from high ionic concentrations in the brush layer causes the brush's polyelectrolyte chains to stretch.[1] Because the distances between the polymer chains average 1.6 nm ($n$ = 55 brush) and 2.2 nm ($n$ = 110 brush), the probe tip with its curvature radius of 10 nm was unable to penetrate the brush layer under the strong stretch condition of the polymer chains just described.

We speculate that, under the multipoint contact condition, the force of detachment of an adsorbed polymer-chain end from the horizontally sliding probe tip is detected as frictional force. This speculation is consistent with the frictional-force behavior of NaPSS brushes shown in Fig. 1. Higher graft density increases the offset of frictional force at zero load because more polymer chains are adsorbed onto the probe tip.

The coefficient of kinetic friction, however, exhibits behavior of the opposite magnitude.

Higher graft density causes the higher ionic concentration and osmotic pressure in the brush layer to stretch the polyelectrolyte chains even more strongly. This condition results in less shrinkage of the polymer chains, thus suppressing the expansion of the contact points against the load. Therefore, lower graft density decreases the coefficient of kinetic friction.

CONCLUSION

We investigated the nanoscale friction properties of NaPSS brushes in water. FFM studies demonstrate that the brushes reduce the coefficient of kinetic friction considerably as compared to a Si wafer and that frictional force is independent of sliding speed. Nanoindentation tests reveal a strong stretching force in the brush layer in water. These results clearly show that the brushes in water stretch strongly against the load with point contact. The contacted points adsorb onto the probe, and the force of their subsequent detachment is detected as frictional force. Therefore, point contact causes offset friction at zero load and reduces the coefficient of kinetic friction.

ACKNOWLEDGMENT

This work was supported by KAKENHI (24560894). We thank Mr. T. Okawa (Omicron NanoTechnology Japan) for his help on performing the nano-indentation measurements.

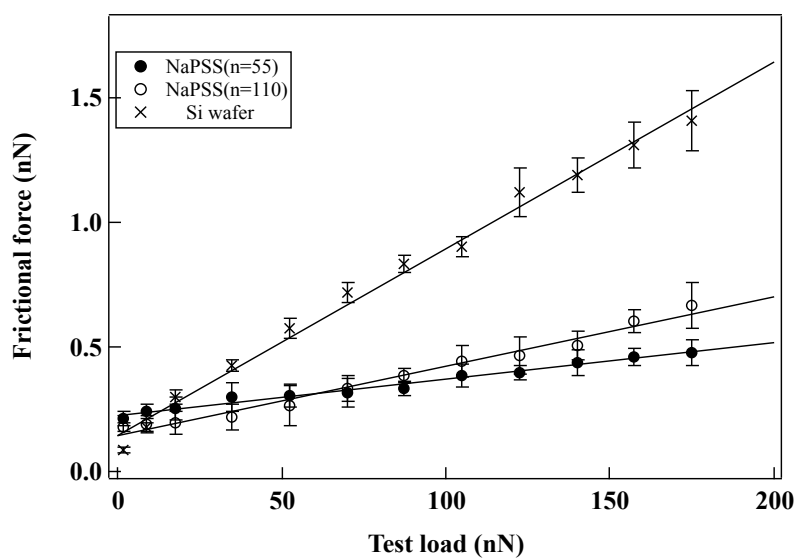

Figure 1. Kinetic frictional force as a function of test load for NaPSS brushes and Si wafer in water. Solid lines were obtained from mean-square analysis.

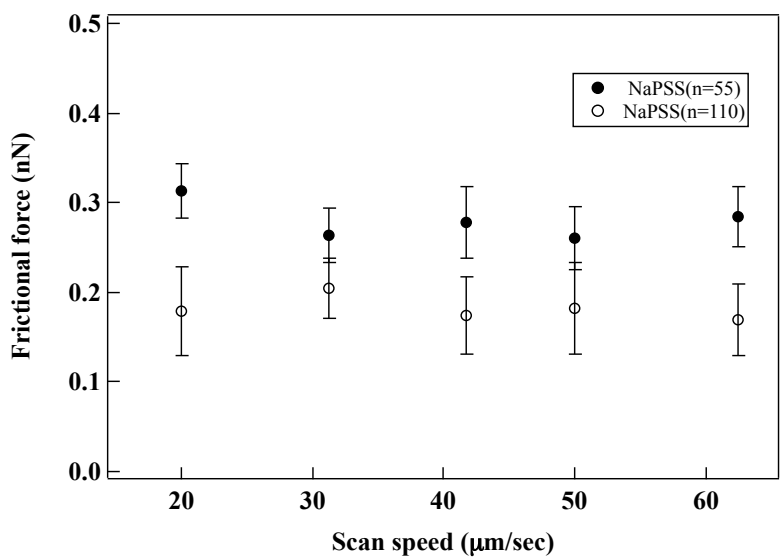

Figure 2. Kinetic frictional force as a function of sliding speed for NaPSS brushes in water.

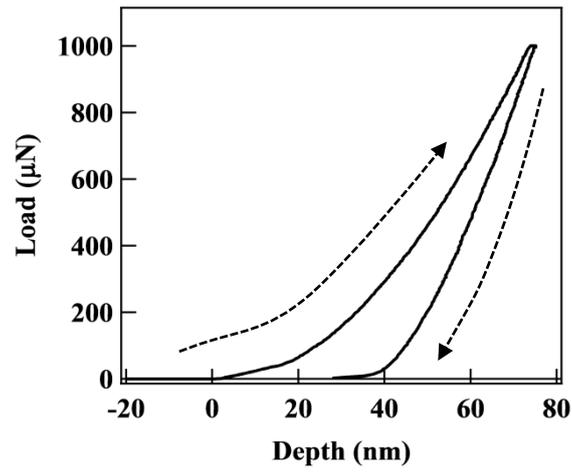

Figure 3. Load as a function of displacement depth during nanoindentation tests of an NaPSS brush ($n$ = 110) in water.

Table 1. Coefficients of kinetic friction for NaPSS brushes and Si wafer measured by FFM in water.

| | |
|---|---|
| NaPSS ($n$ = 55) | $1.5 \times 10^{-3}$ |
| NaPSS ($n$ = 113) | $2.8 \times 10^{-3}$ |
| Si wafer | $7.5 \times 10^{-3}$ |